%% file: Paper_v3.tex
\documentclass[graybox]{svmult}

\usepackage{mathptmx}       % selects Times Roman as basic font
\usepackage{helvet}         % selects Helvetica as sans-serif font
\usepackage{courier}        % selects Courier as typewriter font
\usepackage{type1cm}        % activate if the above 3 fonts are
\usepackage{makeidx}         % allows index generation
\usepackage{graphicx}        % standard LaTeX graphics tool
\usepackage{multicol}        % used for the two-column index
\usepackage[bottom]{footmisc}% places footnotes at page bottom

\makeindex             % used for the subject index
\begin{document}

\title*{Theoretical approaches for nanoscale thermoelectric phenomena}
\author{Carmine Antonio Perroni  and Giuliano Benenti}

\institute{Carmine Antonio Perroni  \at CNR-SPIN  and Dipartimento di Fisica "E. Pancini",\\
Universit\`a degli Studi di Napoli "Federico II",\\
Complesso Monte S. Angelo, via Cinthia, I-80126 Napoli, Italy\\
\email{perroni@fisica.unina.it}
\and Giuliano Benenti \at Center for Nonlinear and Complex Systems, Dipartimento di
Scienza e Alta Tecnologia, Universit\`a degli Studi dell'Insubria, via
Valleggio 11, 22100 Como, Italy\\
Istituto Nazionale di Fisica Nucleare, Sezione di Milano,
via Celoria 16, 20133 Milano, Italy \\
NEST, Istituto Nanoscienze-CNR, I-56126 Pisa, Italy\\
\email{giuliano.benenti@uninsubria.it}}

\maketitle

\abstract{Focus of the chapter is on the theoretical approaches aimed to analyze thermoelectric properties at the nanoscale. We discuss several relevant theoretical approaches  for different set-ups of nano-devices providing estimations of the thermoelectric parameters in the linear and non-linear regime, in particular the thermoelectric figure of merit and the power-efficiency trade-off.  Moreover, we analyze the role of not only electronic, but also of vibrational degrees of freedom. First, nanoscale thermoelectric phenomena are considered in the quantum coherent regime using the Landauer-B\"uttiker method and focusing on effects of energy filtering. Then, we analyze the effects of  many-body couplings between nano-structure degrees of freedom, such as electron-electron and electron-vibration interactions, which can strongly affect the thermoelectric conversion.  In particular, we discuss the enhancement of the thermoelectric figure of merit in  the Coulomb blockade regime for a quantum dot model starting from the master equation for  charge state probabilities and the tunneling rates through the electrodes. Finally, within the non-equilibrium Green function formalism,  we quantify the reduction of the thermoelectric performance in simple models of molecular junctions due to the effects of the electron-vibration coupling and phonon transport  at room temperature.}

\vspace{2pc}
\noindent{\it Keywords}: nanoelectronics, thermoelectricity, quantum dots, molecular junctions

\section{Introduction}

A direct conversion between temperature differences and electric voltages can be achieved through thermoelectric phenomena in solid state systems. In the linear response regime, in addition to a small electric voltage $\Delta V$, a small temperature difference  $\Delta T$ is applied generating an electric current $J_e$ and a heat current $J_h$.
The main transport quantities of the linear regime \cite{DattaBook} are the electrical conductance
$G=J_e/\Delta V $ (at $\Delta T=0$), the thermopower  $S=-\Delta V /\Delta T$ (at $J_e=0$, with $\Delta V$ thermoelectric voltage), the thermal conductance
$K=J_h/\Delta T$ (at $J_e=0$).
The figure of merit $Z$ for the thermoelectric conversion is  defined as
\begin{equation}
Z=\frac{G S^2}{K},
\label{thermoz}
\end{equation}
with the dimension of the inverse of a temperature. In the linear regime, one temperature $T$ typically characterizes the stationary state of the system, therefore one usually considers the dimensionless figure of merit $ZT$.
A very efficient thermoelectric conversion is obtained when $ZT$ is much larger than unity \cite{DattaBook}.

Actual thermoelectric devices work under a finite electric voltage $\Delta V$ and temperature difference $\Delta T=T_1-T_2$, with $T_1$ temperature of the hot reservoir and $T_2$ of the cold one.
Very important quantities are the power $P_{\rm gen}$ delivered to the load and the efficiency $\eta$, which is the ratio between the output power to the load and the heat flow from the hot to the cold reservoir.
Both quantities $P_{\rm gen}$ and $\eta$ can be optimized with varying the parameters of the device \cite{physrep2017,review}. In particular, the efficiency $\eta$ is related to the figure of merit $ZT$ and it can never exceed the Carnot efficiency $\eta_{C}$ defined as
\begin{equation}
\eta_{C}=\frac{\Delta T}{T_1}=1-\frac{T_2}{T_1}.
\label{effic}
\end{equation}
The maximum achievable efficiency $\eta_{\rm max}$
(for a given temperature difference and optimizing over the voltage)
is a monotonous growing function of $ZT$, with
$\eta_{\rm max}=0$ when $ZT=0$ and
$\eta_{\rm max}\to\infty$ when $ZT\to\infty$
(the positivity of entropy production imposes $ZT\ge 0$).

Bulk solid state systems are usually characterized by a small value of the figure of merit $ZT$ \cite{NolasBook}. Even if the charge conductance of metals is high, their thermopower is low. Moreover, in metals, the Wiedemann-Franz law correlates the values of electron thermal and electric conductances limiting the values of $ZT$ \cite{NolasBook}. Actually,  the solid state systems used in bulk thermoelectric devices are doped semiconductors which show intermediate values for both conductance and thermopower \cite{NolasBook}. Moreover, in these semiconductors, phonons, not electrons, typically provide the most important contribution to the thermal conductance.

The use of systems with  reduced dimensionality, such as nanostructures, turns out to be effective to increase the thermoelectric performances \cite{dressel}. Indeed, nanoscale engineering  of electronic and vibrational degrees of freedom allows not only to violate the Wiedemann-Franz law, but also to reduce the phonon thermal conductance \cite{Nanotechnology.26.032001}. High values of $ZT$ have been measured in superlattice thermoelectric devices \cite{venka}, in quantum dot superlattices \cite{harman}, and in semiconductor nanowires \cite{hochbaum}. Moreover, thermopower $S$ gets enhanced in molecular junctions \cite{NNano.8.399}. At the nanoscale, thermoelectric devices can be  analyzed in new linear and non-linear transport regimes, for example the Coulomb-blockade with multi-terminal set-ups \cite{physrep2017}, where effects of electron-electron interactions between charge carriers are relevant \cite{Nanotechnology.26.032001}.
Finally, molecular junctions  \cite{CuevasBook} offer the possibility to study  thermoelectric effects in cases where both electronic and vibrational degrees of freedom cooperate to energy transport  \cite{review}.

This chapter will focus on several relevant theoretical approaches used to interpret the thermoelectric properties of nanoscopic systems. We expose theoretical results in different regimes for electronic and vibrational degrees of freedom, analyzing in particular the absence/presence of many-body interactions onto the nanostructure. For the sake of brevity,  only the main equations for transport properties will be quoted in this chapter. For illustrative purposes, a simple nanostructure model with one electronic level and, in molecular junctions, also one vibrational mode will be frequently used in this chapter.

First, we shall consider thermoelectric phenomena in the quantum coherent regime, investigating the possibility of energy filtering effects.  The devices  can have more than two leads which represent the thermal reservoirs fixing both the electrochemical potentials and temperatures. Linear and non-linear response regime are determined through the Landauer-B\"uttiker method \cite{DattaBook},  hence the transport properties are derived in terms of the transmission function of the nanostructure. Already at this stage, the Wiedemann-Franz law can be violated opening the possibility to tune the nanoscopic resonances and to enhance the thermoelectric performances  \cite{physrep2017,review}.

Then, we will analyze the role played by electron-electron interactions  in enhancing the thermoelectric figure of merit. In particular, we will focus on  the Coulomb-blockade regime for a quantum dot model.  In this regime, the charging energy of the nanostructure represents the most important energy barrier for the transport which, actually, is blockaded at zero bias between source and drain in the limit of small temperature. Tunneling rates through the electrodes are introduced in order to derive the kinetic equations, whose solution provides the charge state probabilities of the Coulomb island.

Finally, we shall consider the effects of room temperature on the thermoelectric performances of nano-devices within the non-equilibrium Green function formalism. Room temperature is the reference for applications, and it favors large values of  thermopower. However, it can activate the vibrational degrees of the leads and the nanostructure influencing the energy transport.  We point out  that both  phonon transport and electron-vibration interactions onto the molecule tend to reduce the thermoelectric efficiency in simple models of molecular junctions at finite temperature.

The chapter is organized as follows. In Sec. II, theoretical results about thermoelectric properties are discussed in the coherent regime. The presence of electron-electron and electron-vibration interactions between nanostructure degrees of freedom is analyzed in Sec. III and IV, respectively, evaluating their effects on thermoelectric conversion. We finish with concluding remarks in Sec. V.

\section{Thermoelectricity in the quantum coherent regime}
\label{sec:landauer}

\begin{figure}[t]
%\sidecaption[t]
\begin{center}
\includegraphics[scale=.38]{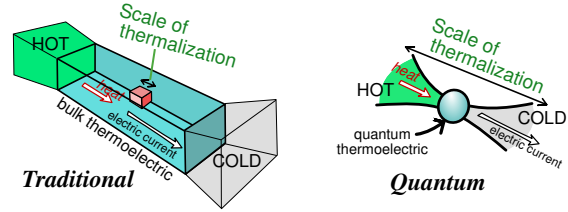}
\end{center}
\caption{A sketch of the qualitative difference between traditional and quantum thermoelectrics. Reprinted with permission from \cite{physrep2017}.}
\label{fig:bulk-vs-nano}       % Give a unique label
\end{figure}

In traditional bulk thermoelectrics (see Fig.~\ref{fig:bulk-vs-nano} left)
the distance on which the electrons relax to a local equilibrum is much smaller
than the system size.
At room temperature, this relaxation length
(typically some tens of nanometres) is usually of the order of the mean free path, since electron scattering is typically dominated by inelastic electron-phonon scattering, which
thermalizes the electrons at the same time as causing electrical resistance by relaxing the electrons' momentum.
Thus, one can treat the electrons inside the thermoelectric structure as being in local thermal equilibrium, with
a local temperature which varies smoothly across the thermoelectric. The system can then be described by Boltzmann transport equations.
In contrast, in nanoscale thermoelectric devices
(see Fig.~\ref{fig:bulk-vs-nano} right), the thermoelectric structure is of size similar or smaller than the lengthscale on which electrons relax to a local equilibrium.
Indeed, at low temperatures (typically less than a Kelvin), electron-electron and electron-phonon interactions are rather weak;
as a result the relaxation length can be many microns (or in some cases even a significant fraction of a millimetre).
Many systems have structures smaller than this, and
one cannot make the approximations necessary to use the standard Boltzmann transport theory.
Consequently the physics becomes much richer, due to quantum interference effects, strong correlations, (quantum) fluctuations and non-equilibrium events that are ubiquitous in all nanoscale devices.
Moreover, transport should be described in terms of conductances rather than conductivities and therefore the figure of merit $ZT$ is not just an intrinsic
material property, but depends on system size, geometry and properties of the contacts with reservoirs.

\subsection{Scattering theory for thermoelectricity}

Landauer-B\"uttiker scattering theory is capable of describing the electrical, thermal and thermoelectric properties of
non-interacting electrons
%(or with interactions described with a
%mean-field Hartree approximation)
in an arbitrary potential (including arbitrary disorder)
in terms of the
probability that the electrons go from one reservoir to another
(of course, it may be challenging to calculate these probabilities
in realistic, complicated nanostructures).
The scattering theory is based on the idea that one can split the situation under consideration into a (small)
scattering region coupled to macroscopic reservoirs of free electrons.  The scattering region should then be such that each electron traverses that region from one reservoir to another without exchanging energy with other particles (electrons, phonons, etc).  Thus, an electron that enters the scattering region with energy $E$ from a given reservoir will be represented by a wave with energy $E$ that bounces around elastically until it escapes into a reservoir.
All inelastic processes that could cause dissipation or decoherence are limited to the reservoirs.
Scattering theory can be a good model even though electrons
interact (repel) each other. This is the case when an electron
feels the average electrostatic effect of the other electrons in a way which is
captured by the mean-field Hartree approximation.
However, it cannot describe (1) the physics of an electron scattering off another one, imparting part of its energy to that electron or (2) the physics of an electron scattering off the lattice (i.e. electron-phonon scattering) and imparting part of its energy to the lattice.
Within scattering theory,  electrons leave the scatter with the same energy that they entered with, each electron only undergoes elastic scattering from the electrostatic potential due to the lattice and the flow of electrons.
For instance, it does not capture the physics of single-electron interaction effects, such as Coulomb blockade
(thermoelectricity in the Coulomb blockade
regime will be described later in
Sec.~\ref{sec:coulomb}).
In this section, we closely follow \cite{physrep2017}. For more details and thorough
discussions on the scattering theory approach, see the textbooks \cite{datta,imry}.

The coupling of the scatterer to each reservoir is written in terms of a set of orthogonal modes in the contact between the scatterer and the reservoir (typically, the transverse modes of a waveguide connecting the system to the reservoirs).
The probability for an electron with energy $E$ to go from mode $m$ of reservoir $\beta$ to mode
$n$ of reservoir $\alpha$ is
\begin{equation}
P_{\alpha n;\beta m}(E)\  =\ \big| {\cal S}_{\alpha n;\beta m}(E) \big|^2,
\label{Eq:scattering-prob}
\end{equation}
where ${\cal S}$ is the scattering matrix.
If we sum this over all modes coupled to reservoirs $\alpha$ and $\beta$, we get the transmission matrix elements
\begin{equation}
{\cal T}_{\alpha \beta}(E) = \sum_{nm} P_{\alpha n;\beta m} (E);
\label{Eq:def-T_ij}
\end{equation}
this can be interpreted as the probability to go from a given mode of reservoir $\beta$ to any mode of reservoir $\alpha$, summed over all modes of reservoir $\beta$.
One has ${\cal T}_{\alpha \beta}(E) \geq 0$ for all $\alpha$, $\beta$, and  $E$.
Moreover, $\sum_\alpha {\cal T}_{\alpha \beta}(E) = N_\beta(E)$
and $\sum_\beta {\cal T}_{\alpha \beta}(E) = N_\alpha(E)$,
where the $\alpha$ and $\beta$ sums are over all reservoirs and
$N_\beta(E)$ is the number of modes in the coupling to reservoir
$\beta$ at energy $E$.

The Landauer-B\"uttiker approach tells us that one can write the charge and heat currents out of reservoir  $\alpha$ in terms of the transmission matrix
elements ${\cal T}_{\alpha \beta}(E)$.  The
charge current $J_{e,\alpha}$ out of reservoir $\alpha$ and into the scatterer is given by counting each electron that crosses the boundary between the scatterer and reservoir $\alpha$.
The number of electrons flowing out of reservoir $\alpha$ and into the scatterer at energy $E$ is proportional to the
number of modes $N_\alpha(E)$ multiplied by the reservoir's occupation at energy $E$, which
is given by the
Fermi function
\begin{equation}
f_\alpha(E) = \left(1+\exp\left[(E - \mu_\alpha)\big/ (k_B T_\alpha) \right] \right)^{-1},
\label{Eq:f}
\end{equation}
where $\mu_\alpha$ and $T_\alpha$  are the electrochemical potential and temperature
of reservoir $\alpha$, and $k_B$ is Boltzmann's constant.
We must also take into account the flow of electrons from the scatterer
into reservoir $\alpha$. The number of electrons that flow into
reservoir $\alpha$ at energy $E$ from reservoir $\beta$ is
proportional to  ${\cal T}_{\alpha \beta}(E)$ multiplied by reservoir
$\beta$'s occupation $f_\beta(E)$.
The total flow of electrons into reservoir $\alpha$ is given
by the sum of this over all $\beta$ (including $\beta=\alpha$).
The electrical current into the scatterer from reservoir $\alpha$ is
then given by the flow of electrons out of the reservoir minus the total flow into it:
\begin{equation}
J_{e,\alpha}  =  \sum_{\beta} \int_{-\infty}^\infty {{\rm d}E \over h} \, e
\, \left[N_\alpha(E)\, \delta_{\alpha\beta} - {\cal T}_{\alpha\beta}(E)  \right] \,  f_\beta (E),
\label{Eq:I-initial}
\end{equation}
where $e$ is the electron charge and $h$ the Planck constant.
We can make the same argument to define the energy current out of reservoir $\alpha$ into the scatterer, except now each electron carries
an amount of energy $E$ instead of the charge $e $.
Hence
\begin{equation}
J_{u,\alpha} = \! \sum_{\beta} \int_{-\infty}^\infty {{\rm d}E \over h} \,  E
\, \left[ N_\alpha(E)\, \delta_{\alpha\beta} - {\cal T}_{\alpha\beta}(E)  \right] \,  f_\beta (E).
\label{Eq:I-energy-initial}
\end{equation}

To construct the equivalent formula for the heat current out of a reservoir,
we must consider the definition of heat in that reservoir.
We take the heat energy in a reservoir's electron gas to be the total energy of the gas
minus the energy which that gas would have in its ground-state at the same electrochemical potential.
As such, the heat energy can be written as a sum over the energy of all electrons, measured from the reservoir's electrochemical potential.  This means electrons above the electrochemical potential
contribute positively to the heat, while those below the electrochemical potential contribute negatively
to the heat.  The latter can be understood as saying that if one removes an electron below
the electrochemical potential, it increases the heat in the reservoir, because one is pushing the system
further from the zero temperature Fermi distribution (in which all states below the electrochemical potential are filled).
Thus, an electron with energy $E$ leaving reservoir $\alpha$ carries an amount of heat, $\Delta Q_\alpha= E -\mu_\alpha$, out of the reservoir.  The formula for heat current is the same as that for energy current, Eq.~(\ref{Eq:I-energy-initial}), but with $(E-\mu_\alpha)$ in place of $E$.
Hence the heat current into the scatterer from reservoir $\alpha$  is
$J_{h,\alpha}=J_{u,\alpha}-(\mu_\alpha/e)J_{e,\alpha}$.
%\begin{equation}
%J_{h,\alpha} = \sum_{\beta} \int_{-\infty}^\infty {{\rm d}E \over h}\, (E\!-\! \mu_\alpha)\,
%\left[ N_\alpha(E)\, \delta_{\alpha\beta} - {\cal T}_{\alpha\beta}(E)  \right]  \, f_\beta (E).
%\label{Eq:J-initial}
%\end{equation}
%Note that $J_{h,\alpha}=J_{u,\alpha}-(\mu_\alpha/e)J_{e,\alpha}$.

Given the above constraints on ${\cal T}_{\alpha\beta}(E)$, we can see that
$\sum_\alpha J_{e,\alpha} = \sum_\alpha J_{u,\alpha} = 0$.
This is nothing but Kirchoff's law of
current conservation for electrical or energy currents.
However, heat currents into the scatterer obey
$\sum_\alpha J_{h,\alpha}=-\sum_\alpha (\mu_\alpha/e) J_{e,\alpha}$.
This means that heat currents are not conserved,
since
the scatterer can be a source or sink for heat.

%It is important to note that the energy current is conserved, but it is not gauge-independent.  That is to say, the value of the energy current, $J_{u,\alpha}$, depends on our choice of the zero of
%energy. This means that the energy current is not of physical relevance,
%although differences in energy currents may be.
%In contrast, even though they are not conserved, the heat currents are gauge-independent.
%Thus they are of physical relevance.

\subsection{Energy filtering}

\begin{figure}[t]
%\sidecaption[t]
\begin{center}
\includegraphics[scale=.22]{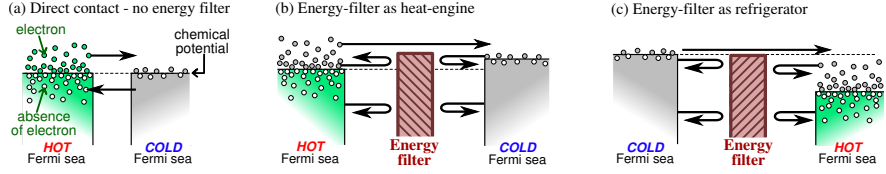}
\end{center}
\caption{Schematic drawing of thermoelectric effects induced by an energy
filter. Reprinted with permission from \cite{physrep2017}.}
\label{fig:energy-filtering}       % Give a unique label
\end{figure}

Thermoelectric effects are present whenever the dynamics of the electrons above the Fermi surface are different from the dynamics of electrons below the Fermi surface.
The simplest example of a thermoelectric effect (captured by scattering theory) is that of an energy filter, schematically illustrated in
Fig.~\ref{fig:energy-filtering}.
In panel (a) we show direct connection between two reservoirs of electrons at different
temperatures but the same electrochemical potential in the absence of any energy filter.
Electrons in occupied (shaded) states want to flow into empty (white) states,
crossing from one reservoir to the other to do so.  The resulting flows are marked by the thick black arrows.
In the absence of an energy-filter there is an heat current but no electrical current (the opposite flows of electrons above and below electrochemical potential cancel each other out).
In panels (b) and (c) we sketch an energy-filter between the hot and cold Fermi seas
which blocks all particle flow below a certain energy.
In (b) we show how to use it as a heat-engine, it generates power because the temperature difference means that electrons flows from a region of lower electrochemical potential (left) to a region of higher electrochemical potential (right).
In (c) we  show how to use it as a refrigerator, using a potential bias
to ensure that electrons above the Fermi sea can flow out of the cold reservoir,
cooling it further.

It turns out that delta-energy filtering (i.e., transport is possible
only in a tiny energy window of width $\delta E\to 0$)
is, for systems described by the scattering theory,
the only mechanism leading to the Carnot efficiency \cite{mahan,linke1,linke2}.
Let us consider two terminals, assuming  $T_1>T_2$, $\mu_1<\mu_2$,
$J_{e,1}> 0$ and $J_{h,1}> 0$. The efficiency for power generation
is then given by
\begin{equation}
\eta=\frac{[(\mu_2-\mu_1)/e] J_{e,1}}{J_{h,1}}=
\frac{(\mu_2-\mu_1)\int_{-\infty}^\infty dE \,{\cal T}(E)\,
[f_1(E)-f_2(E)]}{\int_{-\infty}^\infty
dE \,(E-\mu_1)\, {\cal T}(E)\, [f_1(E)-f_2(E)]},
\end{equation}
where we used the shorthand notation ${\cal T}(E)={\cal T}_{12}(E)$.
When the transmission is possible only within a tiny energy
window around $E=E_\star$, the efficiency reads
\begin{equation}
\eta= \frac{\mu_2-\mu_1}{E_\star-\mu_1}.
\label{eq:etalinke}
\end{equation}
We have $f_1(E_\star)=f_2(E_\star)$,
namely the occupation of states is the same in the two reservoirs
at different temperatures and electrochemical potentials,
when
\begin{equation}
\frac{E_\star-\mu_1}{T_1}=
\frac{E_\star-\mu_2}{T_2}\Rightarrow
E_\star=\frac{\mu_2T_1-\mu_1T_2}{T_1-T_2}.
\end{equation}
Substituting such $E_\star$ into Eq.~(\ref{eq:etalinke}),
we obtain the Carnot efficiency $\eta=\eta_C=1-T_2/T_1$.
Note that Carnot efficiency is obtained in the limit
$J_{e,1}\to 0$, corresponding to \emph{reversible transport}
(zero entropy production) and
\emph{zero output power}.

It is interesting to note that the abstract concept of
delta-energy filtering is illustrated by a single-level quantum dot,
or a single-level molecule. The model is still oversimplified, in that
we make the assumption that the electrons do not interact with each other, so Coulomb blockade effects (which cannot be treated in the scattering theory)
are neglected.
%Most real quantum dots have significant Coulomb blockade effects,
%which we shall discuss in Sec.~\ref{sec:coulomb}.
However, it is none the less instructive to understand the scattering theory for a quantum dot, before
going on to more sophisticated models.
If the dot only has one state at energy $E_\star$, and two reservoirs (each with one mode), which couple
to the dot with strength $w_{ 1}$ and $w_{ 2}$, then
the transmission function has a Lorentzian energy dependence:
\begin{equation}
{\cal T}(E) =  \frac{\Gamma_{ 1} \Gamma_{ 2} }{\left(E-E_\star\right)^2 + \frac{1}{4}\left(\Gamma_{ 1} +\Gamma_{ 2}\right)^2 },
\label{Eq:Briet-Wigner}
\end{equation}
where we define $\Gamma_\alpha=2\pi |w_ \alpha|^2$ for $\alpha\in { 1,2}$.
% such that $\Gamma_\alpha/\hbar$ is the rate at which
%the dot state decays into reservoir $\alpha$.  Thus, we see that the transmission has  a Breit-Wigner form,  with
%$\Gamma_{ 1}+\Gamma_{ 2}$ being the energy-broadening of the dot-level due to the coupling to the reservoirs.
Delta-energy filtering is obtained in the limit $(\Gamma_{ 1}+\Gamma_{ 2})/k_BT\to 0$.

\subsection{Power-efficiency trade-off}

The Carnot limit can be achieved for dissipationless heat
engines. Such ideal machines operate reversibly and infinitely slowly,
and therefore the extracted power vanishes. For any practical purpose it is
therefore crucial to consider the power-efficiency trade-off, in order to
design devices that work at the maximum possible efficiency for a given output power.
This problem was solved, within scattering theory, in~\cite{whitney1,whitney2}.
The first ingredient to derive such bound is the Bekenstein-Pendry
bound~\cite{bekenstein1,bekenstein2,pendry}, which comes from the quantization
of thermal conductance and sets an upper bound on the heat current  through a single transverse mode.
The heat flow out of reservoir $\alpha$ is maximal when that reservoir is coupled to another reservoir at zero temperature (and at the same electrochemical potential) via  a constriction which lets particles flow at all energies:
\begin{equation}
J^{\rm max}_{h,\alpha} =
{\pi^2 \over 6 h}\,  N \, k_B ^2 T_\alpha^2,
\end{equation}
with $N$ number of transverse modes.

The Bekenstein-Pendry upper bound on heat flow places an upper bound
on the power generated by a heat-engine
(since the efficiency is always finite).
A rigourous two-terminal analysis \cite{whitney1,whitney2}
shows that the maximum generated power is
\begin{equation}
P_{\rm gen}^{\rm max} =
 A_0\, {\pi^2 \over h} \, N\,  k_B^2 \big(T_1-T_2\big)^2,
 \label{Eq:Pmax}
\end{equation}
where $A_0 \approx 0.0321$.
This bound is strict in the sense that it is never exceeded, but is achieved by
a system with a transmission function in the form of a step function (i.e. a high-pass filter)
which lets through all particles with $E \geq E_\star$, when one takes
$\Delta\mu\approx 1.146 \,k_B \,\Delta T$, with
$\Delta \mu=\mu_2-\mu_1>0$ and $\Delta T=T_1-T_2>0$.

\begin{figure}[t]
%\sidecaption[t]
\includegraphics[scale=.50]{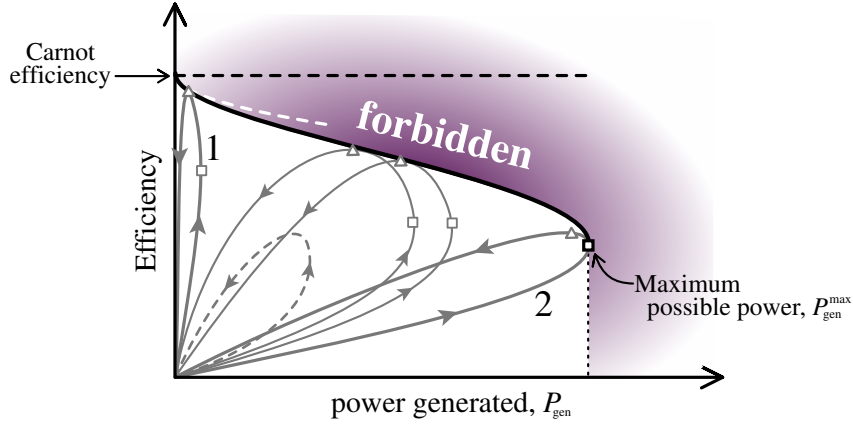}
\caption{Schematic drawing of the
power-efficiency curves (the grey loops) for systems with different transmission functions, at given reservoirs' temperatures $T_1$ and $T_2$.
The thick black curve separates the region accessible by systems with
suitably chosen transmission functions, from the region of efficiencies and powers that no system can achieve.
For small power generation, such curve is approximated by
the dashed white curve given by Eq.~(\ref{Eq:eta-eng-small-Pgen}).
Each loop is formed (in the manner indicated by the arrows) by
increasing the electrochemical potential bias $\Delta\mu$
from zero up to the stopping value, for which the temperature difference
is no longer sufficient to push charges against the bias.
The triangle marks that system's highest efficiency, while the square marks its highest power generation.
Loop 1 is for a system with a narrow transmission function,
which has a low power output, but is capable of
achieving a high efficiency (close to Carnot efficiency).
Loop 2 is for a system with a transmission in the form of a step function,
its maximum efficiency is lower, but its maximum power is much higher.
Reprinted with permission from \cite{physrep2017}.}
\label{fig:bound}       % Give a unique label
\end{figure}

In general, scattering theory implies a bound on  the efficiency at
a given output power $P_{\rm gen}$, which equals the Carnot efficiency at
$P_{\rm gen}=0$, and decays with increasing $P_{\rm gen}$.
A two-terminal calculation find the transmission function
${\cal T}(E)$ that
maximizes the efficiency
of the heat engine for a given output power
(and such bound cannot be overcome for three-terminal devices~\cite{whitney3}).
It turns out that the optimal ${\cal T}$
is a boxcar function, ${\cal T} (E)=1$ for $E_\star<E<{E}_{\star\star}$ and
${\cal T}(E)=0$ otherwise. Here ${E}_{\star\star}$
is determined numerically
by solving the equation ${E}_{\star\star}=\Delta\mu J_{h,1}^\prime/P_{\rm gen}^\prime$, where the prime
indicates the derivative over $\Delta\mu$ for fixed ${\cal T}$
(this equation is transcendental since $J_{h,1}$ and
$P_{\rm gen}$ depend on ${E}_{\star\star}$).
Note that $\Delta\mu$, and consequently $E_\star$, are determined
from the above optimization procedure.
The bound on heat-engine efficiency for a given power output is sketched in
Fig.~\ref{fig:bound}.
At small output power, $P_{\rm gen}/P_{\rm gen}^{\rm max}\ll 1$,
\begin{equation}
\eta \big(P_{\rm gen}\big) \approx  \eta_C
\left(1- 0.478
\sqrt{  {T_2 \over T_1} \ {P_{\rm gen} \over P_{\rm gen}^{\rm max}}}
\right) .
\label{Eq:eta-eng-small-Pgen}
\end{equation}
In the limit ${E}_{\star\star}\to E_\star$, $P\to 0$ and
$\eta\to\eta_C$. In this case, we recover the well-known delta-energy filtering
mechanism to achieve the Carnot efficiency~\cite{mahan, linke1, linke2}.
Note that a similar upper bound can be obtained also for the
efficiency of refrigeration for a given cooling power~\cite{whitney1,whitney2}.
Finally, it is interesting to note that the bound from scattering theory
can be overcome for classical interacting systems~\cite{jiao}.
This result is rooted in the possibility for interacting systems to achieve
the Carnot efficiency at the thermodynamic limit without delta-energy
filtering~\cite{saito2010,jiao2013,carlos2014,shunda2015},
so that large efficiencies can be obtained without greatly reducing power.

\section{Thermoelectricity in the Coulomb blockade regime}
\label{sec:coulomb}

\subsection{Quantum dot model}

A quantum dot (QD) is a paradigmatic model in the investigation of
thermoelectric properties of small devices, since it is characterized
by a spectrum of discrete energy levels. On one side, such system
is ideal to filter the electrons participating in the charge
transport to a narrow energy filter, with (see Sec.~\ref{sec:landauer})
a consequent increase of
the thermoelectric efficiency. On the other side, for
a QD one can also study the effects due to Coulomb interactions among
electrons. Note that in this section we will remain within the Coulomb
blockade regime and we will only consider electron transport,
neglecting any contribution due to phonons.
The discussed results therefore set an upper bound
to the thermoelectric efficiency of the QD, approachable
only in the limit in which suitable strategies to strongly
reduce phonon transport are implemented.
In this section, we shall follow~\cite{erdman}, which
generalizes to multi-terminal setups and to the calculation of
the thermoelectric efficiency (both in the linear response regime
and beyond) the method put forward by~\cite{beenakker1,beenakker2}
(see also~\cite{zianni1,zianni2,zianni3}).

\begin{figure}[t]
\sidecaption[t]
\includegraphics[scale=.65]{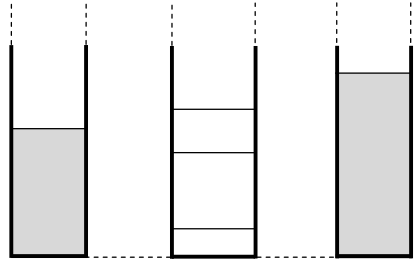}
\caption{Schematic energy representation of a multilevel QD. $E_1$, $E_2$, etc. are the single-electron energy levels of the QD, while $\mu_1$ and $\mu_\alpha$ are the electrochemical potentials relative to reservoir 1 and $\alpha$, respectively. Reprinted with permission from \cite{erdman}.}
\label{fig:multiterminale_energies}       % Give a unique label
\end{figure}

The QD is tunnel-coupled to ${\cal N}$ electron reservoirs, each
characterized by a given temperature $T_\alpha$ and electrochemical
potential $\mu_\alpha$, so that the occupation of the electrons within
reservoir $\alpha$ follows the Fermi distribution $f_\alpha(E)$.
In Fig.~\ref{fig:multiterminale_energies},
$E_p$ (with $p=1,2,\dots$) % labeled in ascending order)
are the QD single-electron energy levels
(these levels can be shifted by means of an applied gate voltage).
Electron-electron (Coulomb) interaction is accounted for by a capacitance
$C$, whose ssociated energy scale is its charging energy $(Ne)^2/2C$,
where $N$ is the number of electrons in the QD.
% and $e$ is the electron charge.
We assume that the QD is weakly coupled to the reservoirs through large
tunneling barriers.
 More precisely, we assume that thermal energy $k_BT$, level spacing and charging energy are much larger than the coupling energy between reservoirs and QD [$\hbar\sum_\alpha\Gamma_\alpha(p)$, where $\Gamma_\alpha(p)$ is the tunneling rate from level $p$ to reservoir $\alpha$, which we assume independent of the number $N$ of electrons inside the dot].
        As a consequence, the charge on the QD is quantized, i.e. each energy level $E_p$ can have either zero or one electron, $n_p=0$ or $n_p=1$ (any degeneracy, like electron spin, can be taken into account counting each level multiple times), and transport occurs due to single-electron tunneling processes ({\it sequential tunneling regime}).
        The electrostatic energy associated with the electrons within the QD is given by $U(N) = E_C N^2$, where $E_C = e^2/2C$, $N = \sum_i n_i$.
%is the total number of electrons within the QD, and $C$ is the capacitance of the QD.

\subsection{Kinetic equations}

        The QD is described by states characterized by a set of occupation numbers $\{n_i\}$ relative to the energy levels.
        The QD changes state whenever a single-electron tunneling process takes place.
       Let $P(\{n_i\})$ denote the non-equilibrium probability for a given state $\{n_i\}$ to occur.
    The single-electron tunneling processes that contribute to changing over time $P(\{n_i\})$ are due to electrons that tunnel from the QD to the reservoirs and vice-versa.
    For an electron exiting the QD, initially with $N$ electrons, from energy level $E_p$ and going into reservoir $\alpha$ at energy $E^{{\rm fin}}$, energy conservation
imposes that
    \begin{equation}
    \label{eq:energy_conservation_fin}
        E_p  +  U(N) = E^{{\rm fin}}(N)+ U(N-1) .
    \end{equation}
    On the contrary, for an electron that tunnels from an initial state in reservoir $\alpha$ at energy $E^{{\rm in}}$ to the level $E_p$ in the QD that initially had $N$ electrons,
%energy conservation imposes that
    \begin{equation}
    \label{eq:energy_conservation_in}
        E^{{\rm in}}(N) + U(N) = E_p + U(N+1) .
    \end{equation}
    $P(\{n_i\})$ can then be determined by the following set of kinetic equations, one for each configuration $\{n_i\}$:
    \begin{eqnarray}
    \label{eq:kinetic_def}
            \frac{\partial}{\partial t} P\left(\{n_i\}\right) =
            -\sum\limits_{p\alpha} \delta_{n_p,0} P\left(\{n_i\}\right) \Gamma_\alpha(p) f_\alpha\left(E^{{\rm in}}(N)\right)
\nonumber \\
            -\sum\limits_{p\alpha} \delta_{n_p,1} P\left(\{n_i\}\right) \Gamma_\alpha(p) \left[ 1-f_\alpha\left(E^{{\rm fin}}(N)\right)\right]
\nonumber \\
            +\sum\limits_{p\alpha} \delta_{n_p,0} P\left(\{n_i\},n_p=1\right) \Gamma_\alpha(p) \left[ 1-f_\alpha\left(E^{{\rm fin}}(N+1) \right)\right]
\nonumber \\
            +\sum\limits_{p\alpha} \delta_{n_p,1} P\left(\{n_i\},n_p=0\right) \Gamma_\alpha(p) f_\alpha\left(E^{{\rm in}}(N-1) \right)  ,
    \end{eqnarray}
    where we have introduced the notation
    \begin{equation}
        P\left(\{n_i\},n_p=1\right) = P\left(\{n_1,\dots,n_{p-1},1,n_{p+1},\dots\}\right)
    \end{equation}
    and
    \begin{equation}
        P\left(\{n_i\},n_p=0\right) = P\left(\{n_1,\dots,n_{p-1},0,n_{p+1},\dots\}\right)
    \end{equation}
    for the QD states. The first term in Eq.~(\ref{eq:kinetic_def}) accounts for the decrease of the probability $P(\{n_i\})$, with the QD initially in the state $\{n_i\}$, due to an electron coming from a reservoir and occupying an empty level in the QD. The rate of electrons coming from reservoir $\alpha$ will be given by a sum over all empty levels $p$ (such that $n_p = 0$) of the tunnel rate $\Gamma_\alpha (p)$, multiplied by the probability of finding the QD in this state, $P\left(\{n_i\}\right)$, and multiplied by the reservoir's occupation $f_\alpha\left(E^{{\rm in}} (N) \right)$ at the correct energy $E^{{\rm in}} (N)$ to tunnel to level $p$.
The second term accounts for the decrease of the probability $P(\{n_i\})$, with the QD initially in the state $\{n_i\}$, due an electron leaving the QD from an occupied level to tunnel into a reservoir.
The third term accounts for the increase of the probability $P(\{n_i\})$ if the QD is in a state with an extra electron in level $p$ with respect to $\{n_i\}$, and if this electron leaves the QD, tunneling to the reservoirs.
The forth term accounts for the increase of the probability $P(\{n_i\})$ if the QD is in a state with a missing electron in level $p$ with respect to $\{n_i\}$, and if this electron enters the QD in level $p$, tunneling from the reservoirs.
    The {\it stationary solution} of the kinetic equations, obtained imposing $\partial P/\partial t = 0$, together with the normalization request
%   \begin{equation}
$
            \sum\nolimits_{\{n_i\}} P\left(\{n_i\}\right) = 1
            \label{eq:normalize_probabilities}
$
%\end{equation}
    provides a complete set of equations that uniquely defines $P\left(\{n_i\}\right)$.
    The sum over $\{n_i\}$ means the sum over $n_i = 0,1$, with $i = 1,2,...$.

    Charge $J_{e,\alpha}$ and energy $J_{u,\alpha}$ currents flowing from reservoir $\alpha$ to the QD can be calculated as the sum of all possible tunneling processes,
    since the QD can be in any state $\{n_i\}$ with probability $P(\{n_i\})$ and an electron can tunnel into or out of any energy level $E_p$.
    For the charge current we have
    \begin{eqnarray}
            J_{e,\alpha} = e \sum\limits_{p=1}^\infty \sum\limits_{\{n_i\}} P(\{n_i\})\Gamma_\alpha(p)\Big\{\delta_{n_p,0}f_\alpha(E^{{\rm in}}(N))
            -\delta_{n_p,1}[1-f_\alpha(E^{{\rm fin}}(N))] \Big\},
    \label{eq:jc_general}
    \end{eqnarray}
%   $e$ being the electronic charge,
while for the energy current we have
    \begin{eqnarray}
                J_{u,\alpha} = \sum\limits_{p=1}^\infty \sum\limits_{\{n_i\}} P(\{n_i\})\Gamma_\alpha(p)\Big\{\delta_{n_p,0}f_\alpha(E^{{\rm in}}(N))
                E^{{\rm in}}(N)
                \nonumber\\
- \delta_{n_p,1}[1-f_\alpha (E^{{\rm fin}}(N))]E^{{\rm fin}}(N)  \Big\},
    \label{eq:ju_general}
    \end{eqnarray}
    $E^{{\rm in}}(N) $ [$E^{{\rm fin}}(N) $] being the energy carried by an electron entering (exiting) the QD.
    The heat currents exiting the reservoirs can then be calculated as $J_{h,\alpha} =  J_{u,\alpha} -({\mu_\alpha}/{e}) J_{e,\alpha}$.

\subsection{Thermoelectric performance in the quantum limit}

Starting from the above expressions for charge and heat currents, one can
obtain in the so-called \emph{quantum limit} and for equidistant levels,
$E_p-E_{p-1}=\Delta E$,
linear-response
analytical expressions for the multi-terminal transport coefficients~\cite{erdman}.
The quantum limit is characterized by having the energy spacing between levels of the QD and the charging energy much bigger than $k_BT$ [while $k_BT\gg \hbar \Gamma_\alpha (p)$].
At low temperatures, the lowest energy levels of the QD are occupied, so that, if there are initially $N-1$ electrons in the QD, electrons can flow mainly through level $p=N$.
        Such process gives the dominant contribution to transport.

In the two-terminal case, one can in particular show~\cite{erdman}
that peaks of both the thermoelectric figure of merit $ZT$ and the
power factor $Q$ [which determines the maximum output power
$P_{{\rm max}}=\frac{1}{4}Q(T_2-T_1)^2$]
are obtained for values of the electrochemical potential
given by $\mu\approx \mu_N\pm 2.40 k_BT$, where
$\mu_N\equiv (N-1) \Delta E+(2N-1)E_C$.
The value $Q^*$ of $Q$ in these points is
$Q^* \approx 0.44 {\gamma k_B}/{T}$
[$\gamma \equiv \Gamma_1\Gamma_2/(\Gamma_1+\Gamma_2)$],
while the value of $ZT$ in the same points is
$ZT^* \approx 0.44\frac{e^{\Delta E/k_BT}}{\left(\Delta E / k_BT\right)^2}$.
This equation shows that
in the limit $\Delta E/ k_BT \rightarrow \infty$,
we have that $ZT^\star \rightarrow \infty$.
For example, for $\Delta E = 6\, k_BT$, we reach $ZT^* \approx 5$;
        for $\Delta E = 10\, k_BT$, we reach $ZT^* \approx 97$, and so on.
        This is consistent with the energy filtering mechanism
discussed in Sec.~\ref{sec:landauer}: a narrow transmission function yields $ZT \rightarrow \infty$. Furthermore, these peaks in $ZT$ correspond to peaks in $Q$, so in these points we can optimize the maximum power
%$P_{{\rm max}}=\frac{1}{4}Q(T_2-T_1)^2$
$P_{{\rm max}}$
and the corresponding efficiency $\eta(P_{{\rm max}})$
(which is a monotonous growing function of $ZT$)
simultaneously.

\begin{figure}[t]
\sidecaption[t]
\includegraphics[scale=.80]{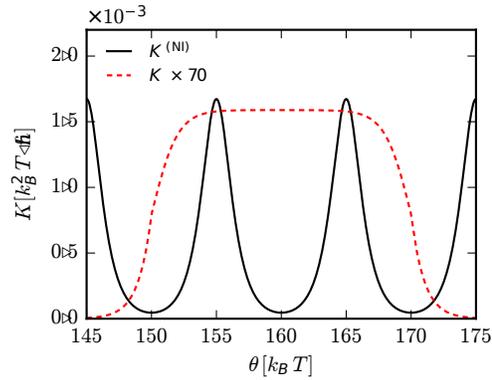}
\caption{Comparison between the thermal conductances
$K^{({\rm NI})}$ and $K$, for the non-interacting and the interacting
case, plotted as a function of the electrochemical potntial $\mu$.
In both cases $\Delta E=10\,k_BT$, and $\hbar\Gamma_1(p) = \hbar\Gamma_2(p) = (1/100)\,k_BT$, while $E_C=50\,k_BT$ in the interacting case and
$E_C=0$ in the non-interacting case. To facilitate comparison,
the interacting thermal conductance has been multiplied by a factor $70$.
Reprinted with permission from \cite{erdman}.}
\label{fig:k_int_noint}       % Give a unique label
\end{figure}

An important question is whether Coulomb interactions may
enhance the thermoelectric
performance of a QD. A positive answer
(in the Coulomb
blockade regime discussed in this section) can be given and
explained~\cite{erdman} by comparing an interacting QD
(with $2E_C > \Delta E\gg k_BT$) with a non-interacting QD
($E_C=0$) that has the same energy spacing $\Delta E$.
Within the linear response quantum limit and for a two-terminal setup,
the electric conductance and the thermopower are not affected
by interactions (neglecting the fine structure
in their dependence on $\mu$, see~\cite{erdman}). On the other hand,
Coulomb interactions dramatically suppress the thermal conductance.
A comparison between the interacting and non-interacting thermal
conductances is plotted in Fig.~\ref{fig:k_int_noint}.
This figure shows a large interaction-induced reduction of the thermal
conductance.
Intuitively, the striking difference between the two models can be explained as follows. If we consider a single energy level QD the thermal conductance vanishes ($K=0$) since $K$ is computed at zero charge current and charge and heat currents are proportional in this case. However, $K$ can be finite when at least two energy levels are available, and gets bigger by increasing the flux of electrons tunneling at different energies. Now, Coulomb interaction produces a correlation between electrons tunneling at different energies. Namely, if one electron enters the QD the electrostatic energy increases by $2E_C$, preventing other electrons from entering the QD at any other energy level. Therefore, until that electron tunnels out of the QD, all other processes are blocked: this is a manifestation of Coulomb blockade. On the contrary, in the non-interacting case all tunneling events are independent. This correlation is thus responsible for suppressing simultaneous tunneling through different energy levels in the interacting case, which results in a suppression of $K$.
As a result of the suppression of $K$ (at constant $G$ and $S$),
Coulomb interactions dramatically enhance $ZT$ (at constant power factor).
For further results on thermoelectric transport in the Coulomb blockade regime,
also beyond linear response, see \cite{erdman}.

\section{Effects of electron-vibration interactions on thermoelectricity}

While so far we focused on low temperatures,
room temperature is the reference for thermoelectric applications.
On one hand an increase of temperature favors large values of  thermopower,
on the other hand it can activate phonon transport in the leads and electron-vibration interactions onto the nanostructure, which can strongly affect thermoelectric efficiency. Temperature effects can become very important in nanoscale systems such as molecular junctions where vibrations represent relevant degrees of freedom for energy exchanges with electrons.

Within the same approximations valid in the Coulomb blockade regime discussed in the previous section, one can include also the effects of electron-vibration interactions. In particular, together with
charging electrostatic energy $U$, there is the additional energy scale $E_P$, the electron-vibration coupling energy, which takes into account the energy barrier due to the interaction of the electrons with the vibrational degrees of freedom. If $E_P$ is the most relevant energy, hence larger than the energy $\hbar \Gamma$ associated to the tunneling onto the nanostructure (thus $\Gamma$ controls the hybridization width of the nanostructure energy levels),  one enters the Franck-Condon blockade regime \cite{PRL.94.206804}. Therefore, one can combine electron-electron and electron-vibration effects into a master equation for electron and phonon number probabilities \cite{PRB.70.195107,PRB.82.045412}  in the presence of both electronic and bosonic (such as phononic) reservoirs.

The rate equations for probabilities take into account only the incoherent charge dynamics in the presence of many-body interactions. One can improve this approach adding co-tunneling effects, that is higher  orders of the perturbative expansion in the tunneling matrix elements \cite{nazarov}.  Otherwise, one should evaluate the density matrix taking into account both diagonal (populations) and off-diagonals (coherences) elements calculating the probability amplitudes between different quantum states of the nanostructure. One way to circumvent some difficulties related to the perturbative expansion in the tunneling processes is to use another method, the non-equilibrium Green's function (NEGF) approach \cite{Haug}. Within this approach, one simply recovers the coherent results in the case where many-body interactions are absent or neglected onto the nanostructure. Indeed, even if interactions are present, the coherent part of the quantum dynamics is preserved. Moreover, NEGF method is able to describe also the blockade transport phenomena induced by many-body interactions.

In the first subsection, we will provide some clues to NEGF approach, in the second one we shall apply this formalism to study electron-vibration effects in simple models of molecular junctions at room temperature.

\subsection{Non-equilibrium Green's Function}

Within NEGF formalism \cite{Haug}, the electrical current $J_{e,\alpha}$ related to the lead $\alpha$ can be expressed in terms of the electron Green functions of the nanostrucutre and the electron self-energies due to the nanostructure-lead coupling:
\begin{equation}
J_{e,\alpha}=\frac{e}{\hbar} \int \frac {d  E}{2 \pi} \,tr_{el} \big\{
 { G}^{>}(E){ \Sigma}^{<}_{\alpha}(E)-
 { G}^{<}(E){\Sigma}^{>}_{\alpha}(E)
\big\},
\label{currentel}
\end{equation}
where $tr_{el} $ is the trace over the electronic degrees of freedom. In equation (\ref{currentel}), the greater electron Green function ${ G}^{>}$ provides the number of available states in the molecule,  ${\Sigma}^{<}_{\alpha}$ the out-tunneling rate of the occupied electronic states in the $\alpha$ lead. Therefore, the first term in equation (\ref{currentel}) provides the current flowing out of the $\alpha$ lead to the nanostructure. Analogously, the lesser electron Green function ${ G}^{<}$ provides the number of occupied states in the nanostrucutre,  ${\Sigma}^{>}_{\alpha}$ the in-tunneling rate of the available electron states in the $\alpha$ lead. Hence, the second term in equation (\ref{currentel}) with the minus sign gives the current flowing from the molecule to the $\alpha$ lead.

The electronic energy current $J_{u,\alpha}$ related to the lead $\alpha$ has an expression similar to the charge current:
\begin{equation}
J_{u,\alpha}=\frac{1}{\hbar} \int \frac{d  E}{2 \pi} \,E \,tr_{el} \big\{
 { G}^{>}(E){\Sigma}^{<}_{\alpha}(E)-
 { G}^{<}(E){\Sigma}^{>}_{\alpha}(E)
\big\}.
\label{currentelen}
\end{equation}

The greater ${G}^{>}$ and the lesser ${G}^{<}$ electron Green functions are related to the retarded   ${ G}^{r}$ and advanced  ${ G}^{a}$ electron Green functions through the Keldysh equation:
\begin{equation}
{G}^{>,<}={ G}^{r}  { \Sigma}_{tot}^{>,<} { G}^{a},
\label{keldysh}
\end{equation}
where ${ \Sigma}_{tot}$ is the total electron self-energy given by the sum of the tunneling contributions ${ \Sigma}_{\alpha}$ and a term ${ \Sigma}_{int}$ due to many-body interactions on the nanostructure:
\begin{equation}
{ \Sigma}^{>,<,r,a}= { \Sigma}_{L}^{>,<,r,a}+ { \Sigma}_{R}^{>,<,r,a}+ { \Sigma}_{int}^{>,<,r,a}.
\label{keldysh2}
\end{equation}

Within NEGF method, equations for fermions and bosons are similar. Indeed,  the phonon energy current $J_{ph,\alpha}$ related to the lead $\alpha$ has  a structure similar to equation (\ref{currentelen}) for electrons.

In the regime of linear response, charge and energy currents allow to determine not only the electron transport quantities, such as the conductance $G$, the thermopower $S$,  and the electron thermal conductance $K^{el}$, but also the  phonon  thermal conductance $K^{ph}$. Hence, the total thermal conductance is $K=K^{el}+K^{ph}$, and the thermoelectric figure of merit becomes  $ZT=G S^2 T /K$.

\subsection{Electron-vibration interactions}

\begin{figure}[t]
%\sidecaption[t]
\begin{center}
\includegraphics[scale=.40]{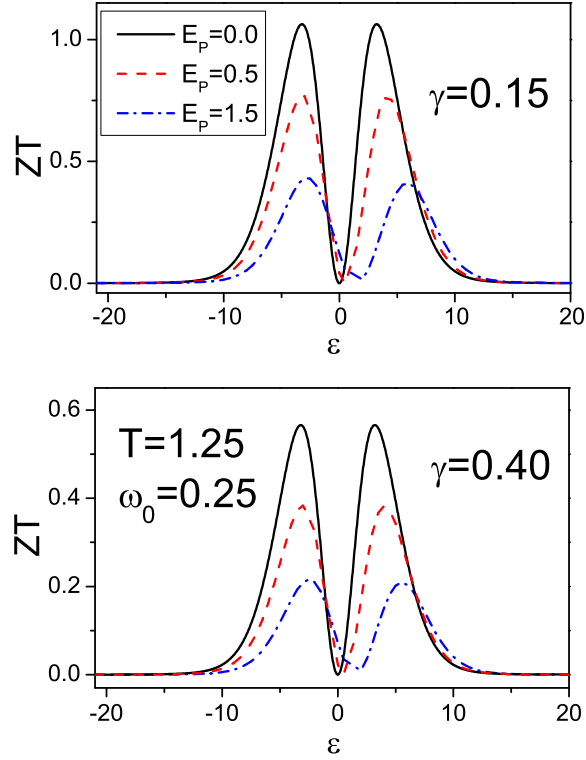}
\end{center}
\caption{Thermoelectric figure of merit $ZT$ at vibrational damping rate $\gamma=0.15$ (in units of $\Gamma$, upper panel) and $\gamma=0.4$ (in units of $\Gamma$, lower panel) as a function of level energy $\epsilon$ (in units of $\hbar \Gamma$) for different values of the electron-vibration coupling $E_P$ (in units of $\hbar \Gamma$). In the plots, $T=1.25$ (in units of $\hbar \Gamma / k_B$) and $\omega_0=0.25$ (in units of $\Gamma$). Reprinted with permission from \cite{Nostro3}.}
\label{GFig8_New}       % Give a unique label
\end{figure}

In this subsection, we analyze the effects of the electron-vibration interaction on the thermoelectric properties starting from spinless  Anderson-Holstein model for molecular junctions \cite{review,CuevasBook}. In particular, we consider a single  electronic level of energy $\epsilon$ coupled to leads with a damping rate $\Gamma$, and a single vibrational mode of low frequency
$\omega_0$ coupled to leads with damping rate  $\gamma$.  For nanodevices with massive molecules, the relevant vibrational degrees of freedom are characterized by low energies, therefore a nonequilibrium adiabatic approach based on the NEGF method has been proposed to study the transport properties  \cite{Nostro1,Nostro2,alberto2,perroni,perroni1,Biggio}. This approach is reliable for $\omega_0 \ll \Gamma$, therefore it is semiclassical. However, it is valid for arbitrary strengths of electron-vibration coupling $E_P$.  The adiabatic approach has been recently employed to study the effects of electron-vibration interaction on the thermoelectric coefficients of junctions with massive molecules, such as fullerene  \cite{Nostro3,Nostro4,Nostro5}.

In Fig.~\ref{GFig8_New}, we report the thermoelectric figure of merit $ZT$ obtained within the adiabatic approach as a function of the level energy $\epsilon$ for different values of the electron-vibration coupling $E_P$ at room temperature $T=1.25$ (in units of  $\hbar \Gamma /k_B$, with $\hbar \Gamma$ of the order of $20$ meV  \cite{Nostro3}). Two values of $\gamma$ are considered: $0.15 \Gamma$ (upper panel of Fig.~\ref{GFig8_New})  and $0.4 \Gamma$ (lower panel of Fig.~\ref{GFig8_New}).  These are somewhat extremal values for $\gamma$ with varying the type and coupling of leads \cite{Nostro3}.  As shown in Fig.~\ref{GFig8_New}, the electron-vibration interaction induces a shift (proportional to the coupling energy $E_P$) of the resonance at higher values of $\epsilon$. With increasing $E_P$, the peak values of  the charge conductance $G$ get reduced, and the minimal and maximal values of the thermopower $S$ are lowered in absolute value. Therefore, the figure of merit $ZT$ gets globally decreased. For the intermediate coupling $E_P=1$, the reduction of the peak in comparison with the value at $E_P=0$ is about twenty-five per cent. For molecules like fullerene, the electron-vibration interaction is estimated to be in the weak to intermediate coupling regime ($E_P<1$).  In any case, the electron-vibration interaction is quite effective in decreasing the thermoelectric performance.  Moreover, from the comparison between upper and lower panel of Fig.~\ref{GFig8_New}, we point out the an increase of the lead phonon-vibration  rate $\gamma$ gives rise to an enhancement of the phonon transport which drastically reduces the figure of merit $ZT$.

%
%\begin{figure}[t]
%\sidecaption[t]
%\includegraphics[scale=.43]{GFig7_New.eps}
%\caption{Electron conductance $G$ (panel (a), in units of $e^2/h$), Seebeck $S$ (panel (b), in units of $k_B/e$), electron thermal conductance $G_K^{el}$
%(panel (c), in units of $k_B \Gamma$), and electronic dimensionless figure of merit $ZT^{el}$ (panel (d), dimensionless) as a function of the level energy $\epsilon$ (in units of $\hbar \Gamma$) for %different values of electron-vibration coupling $E_P$ (in units of $\hbar \Gamma$).  We assume the average chemical potential $\mu=0$ and $T=1.25$ (in units of $\hbar \Gamma /k_B$). In all the %panels, vibrational frequency $\omega_0=0.25$ (in units of $\Gamma$), vibrational damping rate $\gamma=0$ (absence of coupling to phonon leads). Reprinted with permission from \cite{Nostro3}.}
%\label{GFig7_New}       % Give a unique label
%\end{figure}

%
\begin{figure}[t]
%\sidecaption[t]
\begin{center}
\includegraphics[scale=.38]{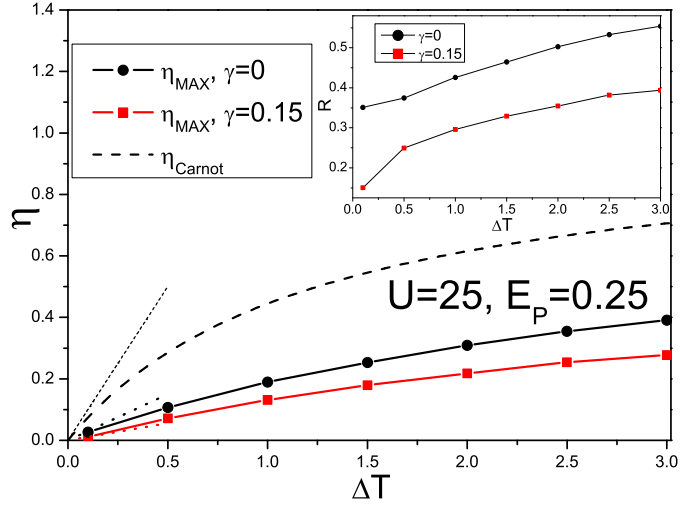}
\end{center}
\caption{ Maximal efficiency $\eta_{\rm max}$ as a function of the temperature difference $\Delta T$ (in units of $ \hbar \Gamma/k_B$)  for $\gamma=0$ (black solid line with circles), and
$\gamma=0.15$ (red solid line with squares, in units of $\Gamma$). In the plot, $U=25 \hbar \Gamma$, $T=1.25 \hbar \Gamma / k_B$ (close to room temperature), $E_P=0.25 \hbar \Gamma$ and $\omega_0=0.25 \Gamma$. The  dashed line reports the Carnot efficiency $\eta_{C}$. The dotted line refers to the slopes in the regime of small $\Delta T$. In the inset, the ratio $R$ between  $\eta_{\rm max}$ and  $\eta_{C}$  for $\gamma=0$ (solid line with black circles), and $\gamma=0.15 \Gamma$ (solid line with red squares). Reprinted with permission from \cite{Nostro4}.}
\label{GFig10_New}       % Give a unique label
\end{figure}

Finally, we focus on the maximal efficiency $\eta_{\rm max}$ as a function of the temperature difference $\Delta T$ in the non-equilibrium Coulomb blockade regime, studying the single level Anderson-Holstein-Hubbard model within the adiabatic approach close to room temperature \cite{Nostro4}. In Fig.~\ref{GFig10_New}, we analyze different values of lead phonon-molecule coupling:
$\gamma=0$  (black solid line with circles) and $\gamma=0.15 \Gamma$ (red solid line with squares). The situation analyzed in the figure corresponds to the case where  $\gamma$ is not large and the electron-vibration coupling is in the weak to intermediate regime. Therefore, the maximal efficiency is not drastically reduced in comparison with the ideal Carnot efficiency $\eta_{C}$ defined in Eq. (\ref{effic}) and reported in Fig.~\ref{GFig10_New} as a dashed line. Indeed, for large $\Delta T$, $\eta_{\rm max}$ for  $\gamma=0$ is about half of the Carnot limit. Moreover, the quantity $\eta_{\rm max}$ for $\gamma=0.15 \Gamma$ is slightly smaller than the maximal efficiency for $\gamma=0$. In the inset of Fig.~\ref{GFig10_New}, we show the ratio $R$ between the maximal efficiency $\eta_{\rm max}$ and the Carnot efficiency  $\eta_{C}$ for different values of $\gamma$ pointing out their different behaviors.  Summarizing, in the case of nano-junctions with weak phonon-molecule and weak to intermediate electron-vibration coupling,  the maximal efficiency gets decreased but it is characterized by a behavior similar to that of ideal performance standards.

\section{Conclusions and perspectives}
In this chapter we have analyzed several important theoretical approaches used for the analysis of thermoelectric phenomena at the nanoscale: Landauer-B\"uttiker method for quantum coherent transport, rate equations for the incoherent Coulomb blockade regime, NEGF for treating electron-vibration couplings in molecular junctions. For each of these methods, we have analyzed advantages and drawbacks, fixing in particular their regime of validity. In the coherent regime, we have discussed energy filtering effects relevant to improve the thermoelectric performances of nano-devices.  In the Coulomb blockade regime, we have pointed out that electron-electron correlations are able to enhance the thermoelectric figure of merit of Coulomb islands. On the other hand, we have stressed that both phonon transport and electron-vibration coupling tend to induce a decrease of thermoelectric conversion in molecular junctions.

In this chapter, we have mainly discussed semi-empirical hamiltonian models for electronic and vibrational degrees of freedom, thus the analysis of {\it ab-initio} theories has been neglected \cite{CuevasBook}.  Moreover, we have focused on the most simple electron-vibration interactions, those Holstein-type with local couplings. In order to improve the description of the transport properties of molecular junctions, it would be interesting to analyze the role of non local electron-vibration interactions \cite{SSH}. Finally, all the approaches discussed in this chapter can be generalized to study the effects induced by time perturbations on the thermoelectric properties in the case without \cite{PRB.83.153417,PRB.93.075136} and with \cite{SR.5.14870,PE.48.36} many-body interactions.

\begin{acknowledgement}
G.B. acknowledges the financial support of the INFN through the project
``QUANTUM''.
\end{acknowledgement}
%
%

\input{referenc_v2}

\end{document}

%% file: referenc_v2.tex
%%%%%%%%%%%%%%%%%%%%%%% referenc.tex %%%%%%%%%%%%%%%%%%%%%%%%%%%%%%
% sample references
% %
% Use this file as a template for your own input.
%
%%%%%%%%%%%%%%%%%%%%%%%% Springer-Verlag %%%%%%%%%%%%%%%%%%%%%%%%%%
%
% BibTeX users please use
% \bibliographystyle{}
% \bibliography{}
%